\documentclass[twocolumn,prb,superscriptaddress,preprintnumbers,amsmath,amssymb,showpacs,floatfix]{revtex4}

\usepackage{graphicx}
\usepackage{dcolumn}
\usepackage{bm}
\usepackage{color}

\newcommand{\led}{$L_{2,3}$}

\newcommand{\fe}{YBaCo$_3$FeO$_7$}
\newcommand{\al}{YBaCo$_3$AlO$_7$}
\newcommand{\os}{YBaCo$_4$O$_7$}

\newcommand{\coz}{Co$^{2+}$}
\newcommand{\cod}{Co$^{3+}$}
\newcommand{\fed}{Fe$^{3+}$}

\begin{document}

\title{Electronic and magnetic properties of the kagom\'e systems YBaCo$_4$O$_7$ and YBaCo$_3$MO$_7$~(M=Al,~Fe)}

\author{N.~Hollmann}
 \affiliation{II. Physikalisches Institut, Universit\"{a}t zu K\"{o}ln,
 Z\"{u}lplicher Str. 77, 50937 K\"{o}ln, Germany}
\author{Z.~Hu}
 \affiliation{II. Physikalisches Institut, Universit\"{a}t zu K\"{o}ln,
 Z\"{u}lplicher Str. 77, 50937 K\"{o}ln, Germany}
\author{M.~Valldor}
 \affiliation{II. Physikalisches Institut, Universit\"{a}t zu K\"{o}ln,
 Z\"{u}lplicher Str. 77, 50937 K\"{o}ln, Germany}
\author{A.~Maignan}
 \affiliation{Laboratoire CRISMAT, UMR 6508 CNRS/ENSICaen,~6~bd~du Mar\'echal,
 F-14050 Caen Cedex 4, France}
\author{A.~Tanaka}
 \affiliation{Department of Quantum Matter, ADSM, Hiroshima University,
 Higashi-Hiroshima 739-8530, Japan}
\author{H.~H.~Hsieh}
 \affiliation{Chung Cheng Institute of Technology,
 National Defense University, Taoyuan 335, Taiwan}
\author{H.-J.~Lin}
 \affiliation{National Synchrotron Radiation Research Center,
 101 Hsin-Ann Road, Hsinchu 30077, Taiwan}
\author{C.~T.~Chen}
 \affiliation{National Synchrotron Radiation Research Center,
 101 Hsin-Ann Road, Hsinchu 30077, Taiwan}
\author{L.~H.~Tjeng}
 \affiliation{II. Physikalisches Institut, Universit\"{a}t zu K\"{o}ln,
 Z\"{u}lplicher Str. 77, 50937 K\"{o}ln, Germany}

\date{\today}

\begin{abstract}
We present a combined experimental and theoretical x-ray
absorption spectroscopy (XAS) study of
the new class of cobaltates
YBaCo$_4$O$_7$ and
YBaCo$_3$MO$_7$ (M= Al, Fe). The focus is on the local electronic
and magnetic properties of the transition metal ions in these
geometrically frustrated kagom\'e compounds. For the mixed valence
cobaltate YBaCo$_4$O$_7$, both the Co$^{2+}$ and Co$^{3+}$ are found
to be in the high spin state. The stability of these
high spin states in tetrahedral coordination is compared with those in the
more studied case of octahedral coordination. For the new compound
YBaCo$_3$FeO$_7$, we find exclusively Co$^{2+}$ and Fe$^{3+}$ as charge states.
\end{abstract}


\maketitle

Cobaltates are currently one of the most investigated classes of
transition metal oxides. They show a rich collection of
interesting physical phenomena, including
superconductivity,\cite{takada03a} giant magneto
resistance,\cite{perez97a} strong thermopower,\cite{terasaki97a}
temperature-driven spin-state and metal-insulator
transitions,\cite{raccah67a} as well as spin-blockade behavior.
\cite{maignan04a, chang09a} These astonishing physical effects are
directly related to the charge, orbital, and spin degrees of
freedom of the Co ions.

Some cobaltates also show a high degree of magnetic
frustration.\cite{maignan00a, rogado02a} Recently, a new class of
cobaltates was introduced\cite{valldor02a} with the compound \os,
which contains kagom\'e layers of tetrahedrally coordinated Co. The strong
antiferromagnetic interactions combined with the geometry in the
lattice in these compounds lead to magnetic frustration. It is
most important to understand the local electronic and magnetic
properties of such frustrated systems in order to model their
magnetic behavior. Yet, very different values are suggested in the
literature concerning the magnetic moments in \os. The first
susceptibility measurements\cite{valldor02a} yielded the large
number of $5.8\mu_B$ per magnetic ion, but later
experiments\cite{chapon06a} gave only $2.2\mu_B$. One neutron
study estimated $\mu_t=3.49\mu_B$ for the ordered moment in the
triangular layer and $\mu_k=2.19\mu_B$ in the kagom\'e
lattice,\cite{chapon06a} while another \cite{soda06a} reported
$\mu_t=1.66\mu_B$ and $\mu_k=1.68\mu_B$, respectively. It is
perhaps a priori also not very clear what moments to expect
theoretically since it is known, for example, that a \cod ion has
the so-called spin state degree of freedom: it can be low spin
(LS, $S$=0, non-magnetic), intermediate spin (IS, $S$=1) or high
spin (HS, $S$=2), depending on the details of the local crystal
field.\cite{haverkort06a,hu04a}

Here we present a study of the local electronic properties of \os,
and its variants \al\ and \fe, using soft x-ray absorption
spectroscopy (XAS) at the Co and Fe $L_{2,3}$ edges. We critically
examine the charge state of the ions as well as their separate
orbital and spin contributions to the magnetic moment.

Single crystals of \fe\ and \al\ were prepared by the floating
zone technique in an image furnace. For the pure cobaltate \os, a
solid-state reaction was used to obtain polycrystalline samples.
All samples have been characterized by x-ray diffraction, magnetic
measurements and electrical resistivity which are published
elsewhere\cite{valldor04a, tsipis05a, tsipis05b, maignan06a, valldor08a}. The materials
are insulators and show highly frustrated magnetic properties. The
XAS experiments were carried out at the Dragon Beamline of the
National Synchrotron Radiation Research Center (NSRRC) in Hsinchu,
Taiwan with an energy resolution of about 0.3 eV. The degree of
linear polarization of the incident light was $\sim 99$\%. Clean sample
areas were obtained by cleaving the crystals \emph{in situ} at
pressures in the low 10$^{-10}$ mbar range. The absorption of the
\led\ edges of Co and Fe was recorded using the total electron
yield method. The oxygen $K$ edge XAS was also measured by using both the 
total electron and fluorescence yield method. The fluorescence detector 
was manufactured by Quantar Technology and is equipped with a grid and 
bias voltages to reject ions and electrons. CoO
and Fe$_2$O$_3$ single crystals were also measured
\emph{simultaneously} to serve as energy reference for the Co and
Fe $L_{2,3}$ edges, respectively.

\begin{figure}[t]
\includegraphics[angle=0,width=8cm]{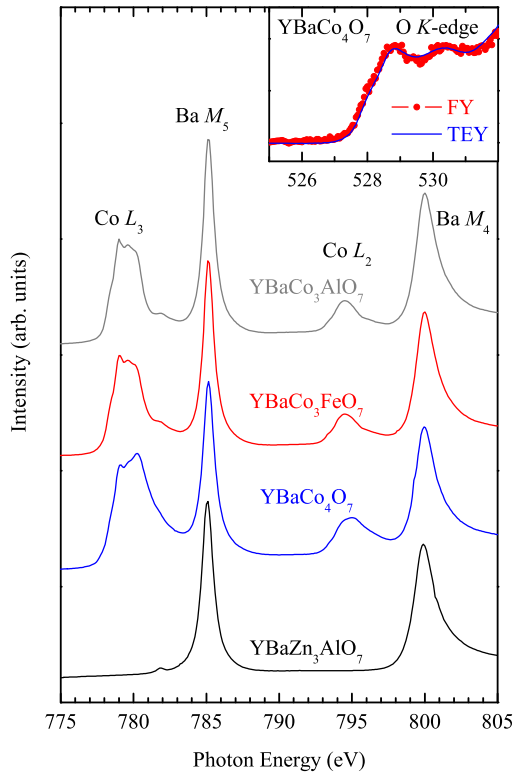}
 \caption[]{(color online) Experimental Co \led\ XAS spectra of \al, \fe, and \os.
The YBaZn$_3$AlO$_7$ spectrum shown at the bottom serves as a
reference for the Ba $M_{4,5}$ edges. The inset depicts the oxygen $K$ XAS spectrum of 
YBaCo$_4$O$_7$ collected by the total electron yield (TEY) and fluorescence yield (FY) methods.
} \label{Ledges}
\end{figure}

Fig.~\ref{Ledges} contains the Co $L_{2,3}$ XAS spectra of \al,
\fe, and \os. The spectra are the result of the dipole-allowed Co
$2p^63d^n\rightarrow 2p^53d^{n+1}$ absorption process. The
spin-orbit coupling of the $2p$ core hole causes the largest
splitting in the spectra: the appearance of the distinct $L_3$
($\approx$780 eV) and $L_2$ ($\approx$795 eV) white lines can be
associated with the $2p_{3/2}$ and $2p_{1/2}$ core hole final
states, respectively. The intense peaks at about 5 eV above the Co
white lines represent the $M_{4,5}$ absorption lines of
barium, i.e. the Ba $3d^{10}4f^0\rightarrow 3d^94f^1$ transitions.
As the cobalt is of interest, we need to subtract the Ba
signal using a reference material. For this the
isostructural compound YBaZn$_3$AlO$_7$ was synthesized. Its Ba $M_{4,5}$ spectrum
is shown at the bottom of Fig.~\ref{Ledges}.
The inset displays the oxygen $K$ XAS spectrum of YBaCo$_4$O$_7$ 
measured in total electron yield and fluorescence yield  
mode. The close similarity of the spectra collected by these two 
methods confirms that our measurements are indeed representative for 
the bulk material and are not plagued by possible surface effects.

\begin{figure}[t]
\includegraphics[angle=0,width=8cm]{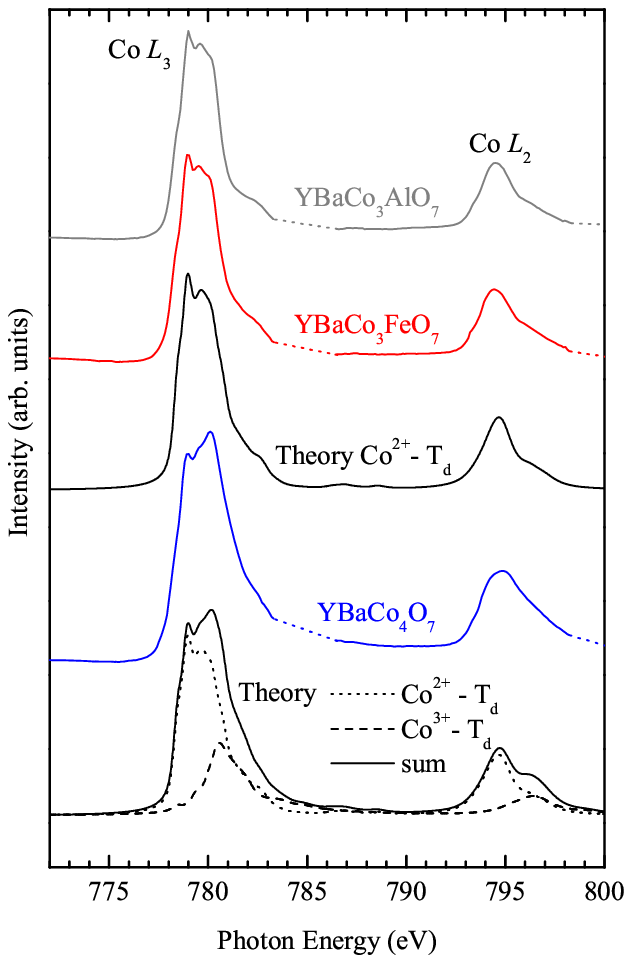}
 \caption[]{(color online) Experimental Co \led\ XAS spectra of \fe, \al, and \os\
after subtraction of the Ba $M_{4,5}$ white lines.  The
theoretical calculation for \coz\ in tetrahedral symmetry is to be
compared with the experimental spectra of \fe\ and \al. The bottom
curve depicts the weighted sum of a calculation for \coz\ and
\cod\ as a simulation for \os.} \label{Co2}
\end{figure}

Fig.~\ref{Co2} shows the Co $L_{2,3}$ edge spectra with the Ba
$M_{4,5}$ signal subtracted. Remnants of the subtraction produce
small glitches at around 785 and 799 eV and are omitted from the
curves. It can be seen that the spectra of \al\ and \fe\ are
practically identical, and that they are different from that of
\os. We have also investigated the possible dependence of the
spectra on the polarization of the light, but we found no
indications for linear dichroism in these \al\ and \fe\
compounds. It is interesting to note that all these spectra are
very different from the ones reported for the more known Co compounds
with CoO$_6$ coordination
like CoO\cite{degroot94a, tanaka94a}, LaCoO$_{3}$\cite{haverkort06a}, LaMn$_{0.5}$Co$_{0.5}$O$_3$\cite{burnus08a}, Ca$_3$Co$_2$O$_6$\cite{burnus08b}, Ca$_3$CoRhO$_6$\cite{burnus08b}, and
La$_{1.5}$Sr$_{0.5}$CoO$_4$\cite{chang09a}.

We first focus on the Co spectrum of \al. From the chemical
formula, for which Al is known to be very stable as a non-magnetic
trivalent ion, one is expecting the Co to be in the divalent
state. The fact that its spectrum is quite different from that of
the divalent Co in CoO, must be sought in the very different local
coordination: tetrahedral in \al\ and octahedral in CoO. To prove
this conjecture, we have carried out a simulation based on the
successful configuration interaction cluster model that includes
the full atomic multiplet theory and the local effects of the
solid.\cite{degroot94a,tanaka94a} It accounts for the
intra-atomic Co $3d$-$3d$ and $2p$-$3d$ Coulomb and exchange
interactions, the atomic $2p$ and $3d$ spin-orbit couplings, the O
$2p$ - Co $3d$ hybridization, and the local crystal field. Here we
set up a CoO$_4$ cluster in $T_d$ symmetry and used a negative
value for the crystal field parameter 10$Dq$ to represent the
splitting between the low two-fold $e$ and the high three-fold
$t_2$ orbitals. The cluster model calculations were done using
XTLS 8.3.\cite{tanaka94a} The parameter values\cite{co2values}
are based on those already known for bulk CoO
\cite{degroot94a, tanaka94a}.
The strength of the O $2p$ - Co $3d$ hybridization was estimated using Harrison's
description\cite{Harrison} for the experimental value of the Co-O bond length.
The 10$Dq$ value was tuned to fit the experimental spectrum.
The simulation result is shown as the
middle curve in Fig.~\ref{Co2}. The model excellently matches the experimental spectrum.
The negative value of 10$Dq = -0.15$ eV, needed in the simulation,
confirms the $T_d$ symmetry of the local coordination of Co$^{2+}$
in \al.
The lack of any polarization dependence in the experiment
is also fully consistent with the completely filled $e$ and
half-filled $t_2$ shell for a Co$^{2+}$ in perfect $T_d$.

\begin{figure}[t]
\includegraphics[angle=0,width=8cm]{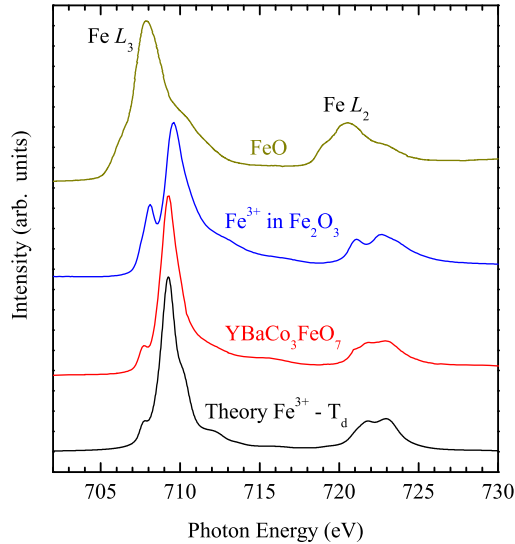}
 \caption[]{(color online) XAS spectrum on the Fe \led\ edges of \fe.
Reference spectra of Fe$^{2+}$ in FeO ($O_h$)and Fe$^{3+}$ in
Fe$_2$O$_3$ ($O_h$) are also given. The spectrum at the bottom is
a simulation for Fe$^{3+}$ in $T_d$ symmetry.} \label{Fe}
\end{figure}

The striking similarity between the \fe\ spectrum with that of \al\ suggests
that all the Co ions are also divalent in \fe.
To fulfill the charge balance, the Fe ions
must be trivalent. To confirm this, we have carried out
XAS measurements on the Fe $L_{2,3}$ edges. The results are shown
in Fig.~\ref{Fe}, along with a reference for Fe$^{2+}$ measured on
FeO\cite{park}, and Fe$^{3+}$ taken from the Fe$_2$O$_3$ single crystal. The Fe
spectrum of \fe\ is quite different from that of FeO. It is also
appreciably dissimilar from that of Fe$_2$O$_3$, although the
energy positions match quite well, strongly suggesting that Fe
in \fe\ could indeed be trivalent. Similar to the case with Co,
these deviations are caused
by the fact that the local coordination is different: $T_d$
in \fe\ vs. $O_h$ in Fe$_2$O$_3$. To clarify this point, we have
carried out simulations for an Fe$^{3+}$ ion in $T_d$ symmetry,
and the result is shown at the bottom of Fig.~\ref{Fe}. The
experimental spectrum can be excellently reproduced. This thus
confirms that the Fe is trivalent. The parameter
values\cite{fevalues} used are reasonable with a negative
10$Dq$ (-0.35 eV) for a $T_d$ local coordination.

We now return to the Co spectrum of \os, which is a mixed
valent compound, with a relation of 3:1 for \coz :\cod\ as
expected from the chemical formula. Here, the spectrum
is constructed as an
incoherent sum of the Co$^{2+}$ and Co$^{3+}$ simulations in $T_d$ symmetry,
weighted in a 3:1 ratio. We note that taking the sum incoherently
is a reasonable approximation because the \os\ compound is an
insulator\cite{tsipis05a, tsipis05b}. The final result
is shown at the bottom of Fig.~\ref{Co2}. The match between
simulated and experimental spectra is very good. For this
simulation we used the Co$^{2+}$ parameters found for the
\al\ compound, and we set the Co$^{3+}$
parameters\cite{co3values} such that the 3d$^{6}$ ion is in the
high spin state (HS, $S$=2).

\begin{figure}[t]
\includegraphics[angle=0,width=8cm]{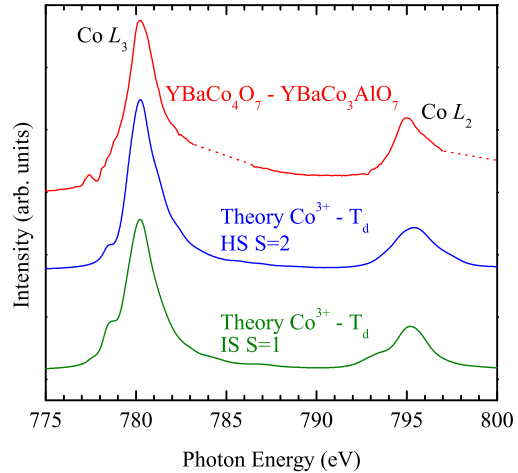}
 \caption[]{(color online)
Experimental Co \led\ XAS spectrum of the Co$^{3+}$ ion in \os\ as
extracted from the subtraction of the \os\ spectrum with the \al\
one in a 4:3 weight ratio. Theoretical simulations for the
Co$^{3+}$ ion in $T_d$ symmetry are also included for the high
spin ($S$=2) and intermediate spin ($S$=1) state
cases.}\label{CoSS}
\end{figure}

As already explained above, a Co$^{3+}$ ion can be HS, IS or LS
depending on the details of the local crystal
field.\cite{haverkort06a, hu04a} To study this aspect more in
detail for the Co$^{3+}$ ion in \os, we first made a subtraction
of the \os\ spectrum with the \al\ spectrum weighted in a 4:3
ratio. The resulting difference spectrum is plotted in
Fig.~\ref{CoSS} and is meant to represent the Co $L_{2,3}$ XAS
spectrum of the Co$^{3+}$ ion in \os. We have carried out
simulations for this Co$^{3+}$ ion in $T_d$ symmetry, and we have
done so for the HS ($e^3t_{2}^3$, $S=2$) and IS ($e^4t_2^2$,
$S$=1) cases, see Fig.~\ref{CoSS}. For the HS state, we have used
10$Dq$ = -0.34 eV, a modest and realistic value close to the -0.15
eV value found for the Co$^{2+}$ ion in the \al\ compound. The
simulation is almost equal to the experimental spectrum. To reach
the IS situation, we have to increase in absolute sense the 10$Dq$
value. As will be explained later, the crossing between the HS and
IS cases is at about 10$Dq=-1.65$ eV, a very large and rather
unphysical number since it is larger than what one could encounter
in an $O_h$ coordination\cite{haverkort06a,ballhausen62a}. For
this particular IS simulation we have used -1.9 eV. The agreement
with the experimental spectrum is reasonable, but there are
distinct features in the simulation, such as the shoulder at 793
eV, which is not present in the experiment. We have not shown
simulations for the LS ($e^4t_2^2$, $S$=0) state, since in $T_d$
symmetry it is not possible to have the two spins of the $t_2$
electrons to be antiparallel. For the LS state to be stabilized,
one would need a very strong local distortion such that the $xy$
orbital in the $t_2$ shell is lowered by at least 0.7 eV with
respect to the $xz$/$yz$ to overcome the local Hund's exchange
interaction. There is no experimental evidence for such strong
distortions. We therefore can conclude that the Co$^{3+}$ ion in
\os\ is most likely in the HS state.

\begin{figure}[t]
\includegraphics[angle=0,width=8.5cm]{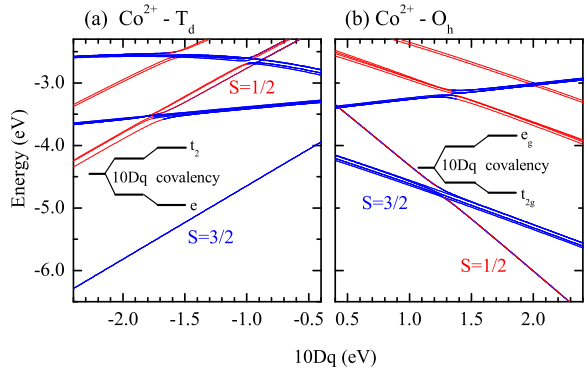}
 \caption[]{Energy level diagrams for \coz\ in a tetrahedral crystal field
 (panel (a), $T_d$, negative 10$Dq$) and octahedral crystal field (panel (b), $O_h$, positive 10$Dq$).
 The full-multiplet calculations include covalency due to O $2p$ - Co $3d$ hybridization.} \label{elevelco2}
\end{figure}

Having established the local electronic structure and having
obtained reasonable estimates for the size of the crystal fields,
we now will look at the consequences for the local magnetic
properties of \coz\ and \cod\ in $T_d$ symmetry with the aid of
full multiplet theory. The ground state of \coz\ in a tetrahedral
crystal field is four-fold spin degenerate, as in first order the
$e$ subshell is completely filled and the $t_2$ subshell
half-filled giving $S=3/2$, leaving no orbital degeneracy. This
would also completely quench the orbital momentum. In second
order, however, the on-site Coulomb and exchange interactions (multiplet effects)
mix a certain amount of $e$
electrons into the $t_2$ shell and restore the orbital momentum
partially. This mixing depends on the relative size of the ligand
field splitting and the spin-orbit coupling and does not
\emph{split} the ground state. But its energy is lowered and the
total magnetic moment is enhanced by the orbital momentum. While
in octahedral complexes the ligand field is often strong enough to
suppress such second order effects, they play a role in
tetrahedral coordination. In the calculation for \coz, the amount
of orbital momentum is 0.5$\mu_B$, increasing of the total magnetic
moment to ~3.5$\mu_B$. In a magnetic susceptibility measurement, the
effective moment at room temperature extracted from the Curie-Weiss law
would be $\mu_{eff}=4.6\mu_B$
(the spin-only value is 3.9$\mu_B$).
Fig.~\ref{elevelco2} (a) shows that in
$T_d$ the lowest quartet $S=3/2$ state is stable over a wide range of
negative 10$Dq$ values. This is to be contrasted to the $O_h$ case,
in which for large enough positive 10$Dq$ the ion converts from a
$S=3/2$ into a $S=1/2$ system, see Fig.~\ref{elevelco2} (b).
Moreover, the $S=3/2$ state in $O_h$ is not a simple
quartet: it is consists of several closely lying sub-levels split
by the spin-orbit interaction giving rise to a non-Curie-Weiss
temperature dependence of the magnetic susceptibility.

\begin{figure}[t]
\includegraphics[angle=0,width=8.5cm]{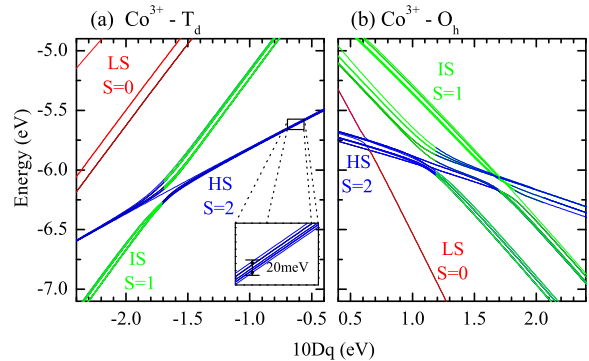}
 \caption[]{Energy level diagrams for \cod\ in a tetrahedral crystal field
 (panel (a), $T_d$, negative 10$Dq$) and octahedral crystal field (panel (b), $O_h$, positive 10$Dq$).
 The full-multiplet calculations include covalency due to O $2p$ - Co $3d$ hybridization.} \label{elevelco3}
\end{figure}

The spin state of \cod\ has been intensively studied for the cubic
and tetragonal local symmetries, but much less so for the
tetrahedral case. Fig.~\ref{elevelco3} (a) shows the energy level
diagram with the energy of the spin states depending on the ionic
$T_d$ crystal field splitting, including hybridization and
spin-orbit coupling. For comparison, a diagram for the $O_h$
coordination is shown on the right panel of Fig.~\ref{elevelco3}
(b). Both panels use the same value for $pd\sigma=-1.53$ eV. A
crossing of the HS state ($t_{2g}^4e_g^2$, $S=2$) to the LS state
($t_{2g}^6$, $S=0$) can be seen at $10Dq\approx 0.65$ eV for $O_h$
symmetry. Such a crossing is also found in $T_d$ symmetry, but at a
significantly higher crystal field energy $10Dq\approx
1.65$ eV, and the ground state turns from a HS state ($e^3t_{2}^3$,
$S=2$) into a IS state ($e^4t_2^2$, $S$=1). The LS ($S$=0) state
is never the ground state in $T_d$ as already explained above. The spin state
found for \os\ is HS, far away from the HS-IS crossing. A
realization of the IS as a ground state for \cod\ in $T_d$ is
indeed very unlikely for two reasons. The first is that the
crossing of the spin states occurs at much higher crystal field
values than found for compounds with Co$^{3+}$ in $O_h$
coordination. The second reason is that it is difficult to
generate such large crystal field values in $T_d$ since the
magnitude of the splitting in $T_d$ is generally smaller than in
$O_h$ for identical bonding lengths between metal and ligand ions,
as it can be seen from point charge calculations that $Dq (T_d) =
-\frac{4}{9} Dq (O_h)$\cite{ballhausen62a}.

The influence of the spin-orbit coupling on the Co$^{3+}$
spin-state has been studied in detail for the octahedral
case\cite{haverkort06a}. It was found that it plays a decisive
role for the magnetic properties of
LaCoO$_3$: the HS state with $S$=2 is split in three sublevels
with $\tilde{J}=1,2,3$, the lowest of which is 3-fold degenerate
and not 5-fold as expected for a $S$=2. By contrast, the IS state
with $S$=1 is split into $\tilde{J}=2,1,0$, resulting in a lowest
state which is 5-fold degenerate and not 3-fold as expected for a
$S$=1. Recognizing these true degeneracies resolved much of the
confusion concerning the interpretation of the magnetic
susceptibility data.\cite{haverkort06a} The situation for a
Co$^{3+}$ ion in $T_d$ is rather different. The stable HS with
$S$=2 state is a 10-fold degenerate $^{5}E$ state if the
spin-orbit coupling effects were absent. The presence of spin-orbit interactions will split
this state into 4 singlets and 3 doublets. The energy splitting however, is
rather small, not much larger than 15-20 meV in total, see inset in Fig.~\ref{elevelco3}.
For temperatures lower than about 20K,
the magnetic susceptibility will be
quite complicated and will depend strongly on the magnitude of the
applied field or effective molecular field in the solid.
In the high temperature regime, the magnetic susceptibility
will be Curie-Weiss like corresponding to a $^{5}E$ ion. The effective
moment at 300K is then estimated to be $\mu_{eff}=5.1\mu_B$, close to the
spin-only value of $4.9\mu_B$.
The total effective moment estimated for \os\ is
$(\frac{3}{4}(4.6\mu_B )^2+\frac{1}{4}(5.1\mu_B )^2 )^{1/2}
=4.7\mu_B$ per Co ion. If these moments would be ferromagnetically
ordered without frustration, the ordered moment is
$\frac{3}{4}3.5\mu_B+\frac{1}{4}4\mu_B=3.6\mu_B$.

To summarize, we have investigated the electronic properties of
\al, \fe, and \os\ in detail. This was done by the use of XAS
spectroscopy and the comparison to configuration interaction
calculations. We have obtained estimates for the crystal fields
and have determined the charge and spin states of the relevant
ions. These findings allowed us to analyze reliably the local
magnetic properties: in $T_d$ symmetry the Co$^{2+}$ ions have
essentially the $S$=3/2 configuration, while the Co$^{3+}$ are in
the high spin $S$=2 state. The reduced ordered moment of Co in
\os\ as seen from neutron scattering must therefore be related to
magnetic frustration. Spin-orbit effects do induce an orbital
moment in the magnetism of \coz\ in all three compounds, but will
not lead to a large magnetic anisotropy as the tetrahedral
structure is almost regular. For Co$^{3+}$, the spin-orbit induced
splitting of the $S$=2 manifold could lead to an intriguing low
temperature magnetic behavior. For the \fe\ compound, we have
determined the Fe charge state to be \fed, meaning that Fe will
act as a magnetic ion with spin-only moment in the compound.
Evidently, this family of cobaltates is a good basis to study
strongly interacting moments on a kagom\'e lattice.

This work is supported by the Deutsche Forschungsgemeinschaft
through SFB 608. NH is also supported by the Bonn-Cologne Graduate School.

\end{document}